\documentstyle{article}
  \newtheorem{theorem}{Theorem}

  \newcommand{\Bbb}{\cal}
  \newcommand{\be}{\begin{equation}}
  \newcommand{\ee}{\end{equation}}
  \newcommand{\bea}{\begin{eqnarray}}
  \newcommand{\eea}{\end{eqnarray}}
  \newcommand{\lb}{\label}
\renewcommand{\a}{\alpha}

\renewcommand{\d}{\delta}
  
\renewcommand{\l}{\lambda}
\renewcommand{\L}{\Lambda}
  \newcommand{\s}{\sigma}
\renewcommand{\S}{\Sigma}

  \newcommand{\bcal}[1]{{\cal #1}}
  
  \newcommand{\calb}{\bcal{B}}
  \newcommand{\calg}{\bcal{G}}
  \newcommand{\ad}{\mbox{\rm ad}}
  \newcommand{\Ad}{\mbox{\rm Ad}}
\renewcommand{\Re}{\,\mbox{\rm Re}\,}
\renewcommand{\Im}{\mbox{\rm Im}\;}
  \newcommand{\ldel}{\langle}
  \newcommand{\rdel}{\rangle}
  \newcommand{\sprod}    [2]{\ldel \,#1\,,\,#2\,\rdel}

\renewcommand{\span}     [1]{\ldel \,#1\,\rdel}
  \newcommand{\lbrac}[2]{[#1,#2]}
  \newcommand{\lbrak}[2]{[#1,#2]}
  
  \newcommand{\supp}{\mathop{\rm supp}}
  \newcommand{\bra}[1]{\ldel\,#1\,|}
  \newcommand{\ket}[1]{|\,#1\,\rdel}
  \newcommand{\braket}[2]{\ldel\,#1\,|\,#2\,\rdel}
  
  \newcommand{\bmrm}[1]{\mbox{\rm #1}}

\begin{document}

\begin{flushright}
Z\"urich University Preprint\\
ZU-TH 38/94
\vskip 0.5\baselineskip
gr-qc/9411058
\end{flushright}
\begin{center}
\vfill
{\LARGE\bf Instability Proof for Einstein-Yang-Mills Solitons and Black
Holes with arbitrary Gauge Groups}
\vfill
{\bf Othmar Brodbeck and Norbert Straumann}
\vskip 0.5cm
Institute for Theoretical Physics\\University of Z\"urich\\
Winterthurerstrasse 190, CH-8057 Z\"urich
\end{center}
\vfill
\begin{quote}
We prove that static, spherically symmetric, asymptotically flat
soliton and black hole solutions of the Einstein-Yang-Mills equations
are unstable for arbitrary gauge groups, at least for the ``generic"
case. This conclusion is derived without explicit knowledge of the
possible equilibrium solutions.
\end{quote}
\vfill

\setcounter{equation}{0}
\renewcommand{\theequation}{\arabic{section}.\arabic{equation}}
\section{Introduction}
In several recent papers \cite{NS1,OB1,OB1.1,OB2}, we have studied
important aspects of the Einstein-Yang-Mills (EYM) system for arbitrary
gauge groups. In particular, we investigated the classification and
properties of spherically symmetric EYM solitons (magnetic structure,
Chern-Simons numbers) and a generalization of the Birkhoff theorem for
the non-Abelian case. We also worked out the generalization of the
first law of black hole physics (Bardeen-Carter-Hawking formula),
allowing for additional Higgs and dilaton fields
\cite{heusler1,heusler2}. For other studies of these and related topics
we refer to \cite{sudarsky,wald1,okai,rogatko}.

In the present paper, we prove that static, spherically symmetric,
asymptotically flat solutions of the EYM equations  are unstable for
any gauge group, if they are ``generic'' (defined in Sec.\ 2). In a
recent letter \cite{OB3}, we have already sketched how we arrived at
this result for solitons. Here we present details of the proof and
extend it to black holes. We discuss also some further mathematical
issues involved.

This general instability was expected since the Bartnik-McKinnon
solutions \cite{bartnik} for the gauge group SU$(2)$ and the related
black hole solutions \cite{volkov,bizon,kunzle1} are unstable
\cite{NS2,NS3,zhou1,zhou2}. A mathematical proof of this expectation
presents, however, quite a challenge, since one cannot rely on any
knowledge of the possible solutions (apart from regularity and boundary
conditions).

Our strategy is based on the study of the pulsation equations,
describing linear radial perturbations of the equilibrium solutions and
involves the following main steps: First we show, that the frequency
spectrum of a class of radial perturbations is determined by a coupled
system of radial, respectively one-dimensional ``Schr\"odinger
equations". Negative parts in the spectrum of the effective Hamiltonian
imply linear instability. With the help of suitably constructed trial
functions, it is then proven, that the spectrum contains always a
negative part (for ``generic'' solutions).

We have recently used a similar procedure to establish the instability
of the gravitating, regular sphaleron solutions of the SU$(2)$
EYM-Higgs system with a SU$(2)$ Higgs doublet \cite{boschung}, which
have been constructed numerically in \cite{greene}. Our results
contain, as a special case, the conclusion of Ref.\ \cite{volkov2} for
the gauge group SU$(2)$. Here, we analyze the regular SU$(2)$ case
further. We show that the effective Hamiltonian for ``sphaleron-like''
perturbations has the form of a ``deuteron'' Hamiltonian.

The paper is organized as follows: In Sec.\ 2 we recall some basic
facts and equations of our previous work \cite{OB1,OB2}, which will be
needed in the present analysis. In Sec.\ 3 we then derive the
linearized perturbation equations for solitons and black holes and
bring them into a convenient, partially decoupled form. The resulting
eigenvalue problem is discussed in Sec.\ 4 and in Sec.\ 5 we show the
existence of unstable perturbations. The ``deuteron'' interpretation
for the unstable modes of a SU$(2)$ soliton is presented in Sec.\ 6. In
the appendix, we elaborate on mathematical issues, related to the
self-adjointness of the effective Hamiltonian and the connection
between the negative part in its spectrum and unstable solutions of the
perturbation equations.
\section{Spherically symmetric EYM fields}
\setcounter{equation}{0}
\renewcommand{\theequation}{\arabic{section}.\arabic{equation}}
We begin with a convenient description of gauge fields with spherical
symmetry (for derivations see \cite{OB1}).

Let us fix a maximal torus $T$ of the gauge group $G$ with
corresponding integral lattice $I$ ( = kernel of the exponential map
restricted to the Lie algebra $LT$ of the torus $T$). In addition, we
choose a basis $S$ of the root system $R$ of real roots. The
corresponding fundamental Weyl chamber
\be
K(S)=\{\, H\in LT \mid \alpha (H)>0 \,\mbox{ \rm for all } \alpha\in S
\,\} \lb{ns1}
\ee
plays an important role in what follows.

To a spherically symmetric gauge field there belongs a canonical
element $H_\lambda\in I\cap\overline{K(S)}$, which characterizes the
corresponding principal bundle $P(M,G)$ over the spacetime manifold
$M$, admitting a SU$(2)$ action. If the configuration is also regular
at the origin, $H_\lambda$ is restricted to a small, finite subset of
$I\cap\overline{K(S)}$, which is described in \cite{OB2}. In the
present discussion, we exclude (for technical reasons) the possibility
that $H_\lambda$ lies on the boundary of the fundamental Weyl chamber.
The term ``generic'' always refers to fields, for which the classifying
element $H_\lambda$ is contained in the {\em open} Weyl chamber
$K(S)$.

The SU$(2)$ action on $P(M,G)$ by bundle automorphisms induces an
action on the base manifold $M$. A SU$(2)$ invariant connection in
$P(M,G)$ defines an invariant connection in each subbundle over a
single orbit of the action on $M$. By Wangs theorem, the induced
connections are described by a linear map $\L\colon
L\bmrm{SU}(2)\rightarrow LG$, which depends locally smoothly on the
orbit and satisfies
\be
\L_1=[\L_2,\L_3]\:, \qquad
\L_2=[\L_3,\L_1]\:, \qquad
\L_3 =-H_\l/4\pi\:,
\lb{ns2}
\ee
where $\L_k:=\L(\tau_k)$ and $2i\tau_k$ are the Pauli matrices.
These equations imply that $\L_+:=\L_1+i\L_2$ lies in the following
direct sum of root spaces $L_\alpha$ of $LG_{\Bbb C}$:
\be
\renewcommand{\arraystretch}{1.3}
\begin{array}{cc}
{\displaystyle
\Lambda_+ \in  \bigoplus_{\alpha\in S(\l)}\;L_\a}\:,
\\
{\displaystyle
S(\l):=\{\,\a\in R_+\mid\a(H_\l)=2\,\}}\:,
\lb{ns3}
\end{array}
\renewcommand{\arraystretch}{1}
\ee
where $R_+$ denotes the set of positive roots in $R$ (relative to the
basis S). In the generic case $S(\l)$ turns out to be a basis of a root
system contained in $R$ (see appendix A of Ref.\ \cite{OB2}).

The $LG$-valued functions $\L_\pm$ on the orbit space determine part of
the connection on $P(M,G)$. Before we give a parametrization of the YM
potential in a convenient gauge, we fix our conventions in
parametrizing the Lorentz metric on $M$ and introduce some further
notation. We use standard Schwarzschild-like coordinates and set
\be
ds^2= -NS^2 dt^2 + N^{-1} dr^2 + r^2({d\vartheta}^2 +
\sin^2\!\vartheta\,{d\varphi}^2)\:,
\lb{ns5}
\ee
where the metric functions $N=:1-2m/r$ and $S$ depend only on $r$ and
$t$.

A suitably normalized  $\Ad (G)$-invariant scalar product on $LG$ will
be denoted by $\ldel\,\cdot\,,\cdot\,\rdel$. We use the same symbol for
the hermetian extension to $LG_{\Bbb C}$ (linear in the second
argument), and $|\cdot|$ means the corresponding norm. Note that the
original $\Ad(G)$-invariance extends on $LG_{\Bbb C}$ to
\be
\sprod{X}{\lbrak{Z}{Y}}+\sprod{\,\lbrak{c(Z)}{X}}{Y}=0\:,
\lb{ns10}
\ee
where $c$ is the conjugation in $LG_{\Bbb C}$.

In Ref.\ \cite{OB1}, it is shown the gauge potential $A$ can be chosen
to have the form
\be
A=\tilde{A}+\hat{A}\:,
\lb{ns6}
\ee
with
\be
\hat{A}={\Lambda_2
\,d\vartheta}+{(\Lambda_3\cos\vartheta-\Lambda_1\sin\vartheta)\,d\varphi}
\lb{ns7}
\ee
and
\be
\tilde{A}={\tilde{\cal A}}\,dt+{\tilde{\cal B}}\,dr\:,
\lb{ns8}
\ee
where ${\tilde{\cal A}}$ and ${\tilde{\cal B}}$ commute with $H_\l$
(i.e. with $\L_3$). Since $H_\l$ is assumed to be generic, its
centralizer is the infinitesimal torus $LT$. Hence, ${\tilde{\cal A}}$
and ${\tilde{\cal B}}$ are $LT$-valued and $\tilde{A}$ is thus
abelian.

For the example of the gauge group SU$(2)$, $H_\l$ is an integer
multiple of $4\pi\,\tau_3$: $H_\l=4\pi k\,\tau_3$ with $k\in{\Bbb Z}$,
and the only solutions of (\ref{ns2}) are $\L_1=\L_2=0$,
$\L_3=k\,\tau_3$ and
\be
\L_1=w\,\tau_1+\tilde{w}\,\tau_2\:,\qquad
\L_2=\mp\tilde{w}\,\tau_1\pm w\,\tau_2\:,\qquad
\L_3=\pm\tau_3\:.
\lb{ns9}
\ee

The gauge potential $A$ contains a ``trivial'', abelian part, which
decouples from the EYM equations. To demonstrate this, let us first
construct a convenient decomposition of $LT$. For a given potential we
restrict the sum in Eq.\ (\ref{ns3}) to the smallest subset $\S$ of
$S(\l)$ for which
\be
\Lambda_+ \in  \bigoplus_{\a\in\S\subset S(\l)}\;L_\a\:.
\lb{E1}
\ee
Since every rootspace $L_\a$ is $\Ad (T)$-invariant and since the
residual gauge group of the potential $A$ is just the torus $T$, the
subset $\S$ is unique and depends only on the invariant connection.
With the help of $\S$ we now split $LT$:
\be
LT=\span{\S} \oplus \span{\S}^{\perp},
\lb{E2}
\ee
where $\span{\S}$ denotes the linear span of $\S$. The decomposition
(\ref{E2}) is {\em independent} of the chosen $\Ad (G)$-invariant
scalar product \cite{OB2} and satisfies
\be
\lbrac{\span{\S}^\perp}{\L_+}=0\;.
\lb{E3}
\ee
This property motivates to set
\be
\begin{array}{ccl}
\tilde{\cal A}&=& a+{\cal A}\;,\\
\tilde{\cal B}&=& b+{\cal B}\;,\\
	    \L_3&=&\L_{3\perp}+\L_{3\parallel}\;,
\end{array}
\lb{E4}
\ee
with
$a$, $b$, $\L_{3\perp}$ $\in$ $\span{\S}^{\perp}$ and ${\cal A}$,
${\cal B}$, $\L_{3\parallel}$ $\in$ $\span{\S}$.
For our instability proof we adopt the following (mixed) gauge:
\be
{\cal A}\equiv0\;,\qquad
       b\equiv0\;,\qquad
       {\displaystyle\lim_{r\rightarrow\infty}}a=0\;.
\lb{E5}
\ee

If we now insert the parametrizations (\ref{ns5}), (\ref{ns6}) --
(\ref{ns8}), (\ref{E4}), (\ref{E5}) into the EYM equations, we obtain a
system of partial differential equations for the metric functions $N$,
$S$ and the YM amplitudes $\L_\pm$, $\calb$. As noted above and as
Eq.\ (\ref{E3}) indicates, the equation for $a$ decouples. Specializing
the results of \cite{OB1} (and using slightly different notation), they
read as follows:

The Einstein equations give two constraint equations for the $r$
derivative (denoted by a dash) and the $t$ derivative (denoted by a
dot) of $m$
\be
m'= {\kappa\over 2}
\Bigl\{
NG+r^2 p_\theta
\Bigr\}\:,
\qquad
\dot m={\kappa\over 2}NH\:,
\lb{ns11}
\ee
($\kappa:=8\pi G$), and the $(rr)$-equation reduces to
\be
{S'\over S}={\kappa\over r}G\:,
\lb{ns12}
\ee
where
\addtolength{\jot}{5pt}
\begin{eqnarray}
G &=& \frac{1}{2}
\biggl\{\,
(NS){\vphantom{|\L_+|}}^{-2}\,|\dot\L_+|{\vphantom{|\L_+|}}^2
	  +|\,\L_{+}'+[\bcal{B},\L_+]\,|{\vphantom{|\L_+|}}^2
\biggr\}\:,\lb{ns13} \\
H &=& \mbox{Re}\,
\bigl\ldel\,
\dot\L_{+}\,,\,\L_{+}'+[\bcal{B},\L_+]\,
\bigr\rdel\:,\lb{ns14}\\
p_\theta &=& {1\over 2r^{4}}
\biggl\{
 |{\hat   {\cal F}}_\parallel|{\vphantom{|{\cal F}}}^2
+|{\check {\cal F}}_\parallel|{\vphantom{|{\cal F}}}^2
+|P_\perp|^2
+|Q_\perp|^2
\biggr\}\lb{ns15}
\end{eqnarray}
\addtolength{\jot}{-5pt}

\noindent with
\be
{\hat  {\cal F}}_\parallel =
\frac{i}{2}\,[\L_+,\L_-]-\L_{3\parallel}\:,
\qquad
{\check{\cal F}}_\parallel = \frac{r^2}{S}\,\dot\bcal{B}
\lb{ns16}
\ee
and
\be
P_\perp=\L_{3\perp}\:,\qquad
Q_\perp=-\frac{\,r^2}{S}a'\:.
\lb{E6}
\ee

The YM equations decompose into

\vspace{.5ex}
\be
\frac{2}{NS}
\biggl(
\frac{r^2}{S}\,\dot\bcal{B}
\biggr)\!{\vphantom{\biggr)}}^{\textstyle\cdot}
+ \Bigl[\,\L_+\,,\,\L_-'+[\bcal{B},\L_-]\,\Bigr]
+ \Bigl[\,\L_-\,,\,\L_+'+[\bcal{B},\L_+]\,\Bigr]=0\:,
\lb{ns17}
\ee

\vspace{-2.25ex}
\begin{eqnarray}
\frac{1}{S}
\biggl(
\frac{1}{NS}\,\dot\L_+
\biggr)\!{\vphantom{\biggr)}}^{\textstyle\cdot}
\!\!\!&-&\!\!\!
\frac{1}{S}
\biggl(NS\Bigl\{\,
\L_+'+[\bcal{B},\L_+]
\,\Bigr\}\biggr)'
\nonumber\\
\!\!\!&-&\!\!\!
N
\Bigl[\,\bcal{B}\,,\,\L_+'+[\bcal{B},\L_+]\,\Bigr]
+\frac{i}{r^2}\,[{\hat {\cal F}}_\parallel,\L_+]=0\:,
\lb{ns18}
\end{eqnarray}

\vspace{-1.5ex}
\be
2
\biggl(
\frac{r^2}{S}\,\dot\bcal{B}
{\biggr)}'
+\frac{1}{NS}
\biggl\{\,
[\L_+,\dot\L_-]+[\L_-,\dot\L_+]
\,\biggr\}
=0\:.
\lb{ns19}
\ee

\vspace{3ex}\noindent
The abelian electric part of the potential satisfies
\be
Q_\perp=-\frac{\,r^2}{S}a'=\mbox{constant}\quad(\in\span{\S}^\perp)
\lb{E7}
\ee
and hence decouples.

Eqn. (\ref{ns19}) is the Gauss constraint. For static solutions all
time derivatives disappear, $\calb$ can be gauged away and the basic
equations simplify considerably. (For the Bartnik-McKinnon solutions
$\L$ is of the form (\ref{ns9}) with $\tilde w=0$,
$\L_{3\parallel}=\L_3=\tau_3$ and $\tilde A=0$ in (\ref{ns6}).)
\section{Perturbation equations}
\setcounter{equation}{0}
\renewcommand{\theequation}{\arabic{section}.\arabic{equation}}
In this section we study time-dependent perturbations of a given
static, asymptotically flat solution of the coupled EYM equations
(\ref{ns11}), (\ref{ns12}), (\ref{ns17}) -- (\ref{E7}). Regular
solutions are ``purely magnetic'' ($\tilde A=0$ in (\ref{ns6})) with
vanishing YM charge ($P_\perp$=$Q_\perp$=$0$ and
$\lim_{r\rightarrow\infty}{\hat{\cal F}}_\parallel$=$0$).
Unfortunately, this is not yet proven with satisfactory weak fall-off
conditions, but there is strong evidence for this (see
\cite{OB2,kunzle2} for partial results.) The perturbation equations we
derive hold also for black holes, if their gauge potentials $A$ have
the form
\be
A=a\,dt+\hat{A}
\lb{F1}
\ee
with
\be
a(r)=Q_\perp\int_r^\infty\frac{S}{y^2}\,dy
\ee
for a constant vector $Q_\perp$ in $\span{\S}^\perp$ (i.e. ${\cal
A}={\cal B}=0$ in Eq.\ (\ref{E4})). We call such gauge fields
``essentially magnetic''.

 From now on $\L_\pm$, $N$, $S$, etc. refer to an essentially magnetic
equilibrium solution and time-dependent perturbations are denoted by
$\d\L_\pm,\d\bcal{B}$, etc.. All basic equations are linearized around
the equilibrium solution. In order to decouple the perturbation $\d a$,
we impose the additional constraint $\d Q_\perp$=$0$.

First, we linearize the right hand sides of the Einstein equations
(\ref{ns11}) and (\ref{ns12}). Since $\calb$ and $\dot\L_\pm$ vanish
for the equilibrium solution, the first order variation of the source
$G$ is
\be
\d G=\Re\sprod{\L'_+}{\d\L'_+}
   -\Re\sprod{\L'_+}{\lbrak{\L_+}{\d\calb\,}}\:.
\ee
Here, the last term vanishes, because the property (\ref{ns10}) of the
scalar product implies
\be
-2\Re\sprod{\L'_+}{\lbrak{\L_+}{\d\calb\,}}
=
\sprod{\,\lbrak{\L_+}{\L'_-}+\lbrak{\L_-}{\L'_+}}{\d\calb}\:,
\ee
and the YM equation (\ref{ns17}) for the equilibrium solution shows
that
\be
\lbrak{\L_+}{\L'_-}+\lbrak{\L_-}{\L'_+}=0\:.
\lb{ns20}
\ee
Thus,
\be
\d G=\Re\sprod{\L'_+}{\d\L'_+}\:.
\lb{ns21}
\ee
The only first order variation for $p_\theta$ comes from $\d |\hat
{\cal F}_\parallel|{\vphantom{|{\cal F}}}^2=2\sprod{\hat{\cal
F}_\parallel}{\d\hat{\cal F}_\parallel}$. Using
\be
\d\hat{\cal F}_\parallel = \frac{i}{2}\,[\L_+,\d\L_-]
		-\frac{i}{2}\,[\L_-,\d\L_+]
\lb{ns22}
\ee
(see Eq.\ (\ref{ns16})), we have
\be
\d p_\theta = {1\over r^{4}}\Re\sprod{i\,\lbrac{\hat{\cal
F}_\parallel}{\L_+}}{\d\L_+}\:.
\lb{ns23}
\ee

Now we can work out the variation of the first Einstein equation in
(\ref{ns11}). With (\ref{ns21}), (\ref{ns23}) and (\ref{ns12}) for the
equilibrium solution, we find
\be
\d m'=
-\,{S'\over S}\,\d m
+{\kappa\over 2}\,
\biggl\{
   N\Re\sprod{\L'_+}{\d\L'_+}
+ \Re\sprod{\,{i\over r^{2}}\lbrac{\hat{\cal
F}_\parallel}{\L_+}}{\d\L_+}
\biggr\}\:.
\ee
For the commutator in the last term we use the unperturbed YM equation
(\ref{ns18}), i.e.
\be
\frac{i}{r^2}\,[\hat {\cal F}_\parallel,\L_+]=
N\frac{S'}{S}\L'_+ + N'\L'_+ + N\L''_+\:,
\lb{ns24}
\ee
whence
\be
\d m'=
-\,{S'\over S}\,\d m +
{S'\over S}\biggl\{\,
{\kappa\over 2} N\Re\sprod{\L'_+}{\d\L_+}\,\biggr\}
+
\biggl\{\,
{\kappa\over 2} N\Re\sprod{\L'_+}{\d\L_+}
\,\biggr\}'
\ee
or
\be
(\d m\,S)'=
\biggl\{\,
{\kappa\over 2} NS\Re\sprod{\L'_+}{\d\L_+}
\,\biggr\}'\:.
\ee
Therefore, $\d m$ must be of the form
\be
\d m=
{\kappa\over 2} N\Re\sprod{\L'_+}{\d\L_+}+{f(t)\over S}\:,
\lb{ns25}
\ee
where $f(t)$ is a function of $t$ alone. This function is determined by
considering the variation of the second Einstein equation in
(\ref{ns11}), which reads
\be
\d \dot{m}=
{\kappa\over 2} N\Re\sprod{\L'_+}{\d\dot{\L}_+}\:.
\ee
Thus, we have also
\be
\d m=
{\kappa\over 2} N\Re\sprod{\L'_+}{\d\L_+}+g(r)\:,
\lb{ns26}
\ee
with a function $g(r)$ of $r$ alone. By comparing (\ref{ns25}) and
(\ref{ns26}), we arrive at the remarkably simple result
\be
\d m=
{\kappa\over 2} N\Re\sprod{\L'_+}{\d\L_+}\:,
\lb{ns27}
\ee
which generalizes an observation already made in \cite{NS2}.

The variation of the Einstein equation (\ref{ns12}) is immediately
obtained with (\ref{ns21})
\be
\d
\biggl(
{S' \over S}
\biggr)
={\kappa\over r} N\Re\sprod{\L'_+}{\d\L'_+}\:.
\lb{ns28}
\ee

Before also linearizing the YM equations, we introduce a suitable
decomposition of $\L_+$ and $\d\L_+$. To do so, we choose a base
element ${e_\a}$ of the root spaces $L_\a$ and expand the unperturbed
$\L_+$ as well as its perturbation $\d\L_+$:
\be
  \L_+=\sum_{\a\in\S}\,   w^\a\, e_\a\:,\qquad
\d\L_+=\sum_{\a\in\S}\,\d w^\a\, e_\a\:.
\lb{ns29}
\ee
Then, we have
\be
\d\L_\pm=\d X_\pm \pm i\d Y_\pm
\lb{ns30}
\ee
with
\be
\d X_+=\sum_{\a\in\S}\Re(\d w^\a)\, e_\a\:,\qquad
\d Y_+=\sum_{\a\in\S}\Im(\d w^\a)\, e_\a
\lb{ns31}
\ee
and the corresponding expansion for $\d X_-$ and $\d Y_-$ with $e_\a$
replaced by  $c(e_\a)\in L_{-\a}$, because $\d\L_-=c(\d\L_+)$ and thus
\be
\d X_-=c(\d X_+)\:,\qquad \d Y_-=c(\d Y_+)\:.
\lb{ns32}
\ee
We call $\d X_\pm$,  $\d Y_\pm$ the ``real'' (or ``gravitational'') and
``imaginary'' (or ``spha\-le\-ron-like'') parts of the perturbations  $\d
\L_\pm$. It was shown in \cite{OB2} that the unperturbed  $\L_+$ can be
chosen to have only a {\em\/real\/} part.

This decomposition will lead to a significant decoupling of the
perturbation equations. Note in particular, that the variations $\d m$
and $\d p_\theta$ in (\ref{ns23}) and (\ref{ns27}) depend only on the
real part $\d X_+$:
\addtolength{\jot}{5pt}
\begin{eqnarray}
\d m       &=&
{\kappa\over 2} N\sprod{\L'_+}{\d X_+}\:,
\lb{ns33}\\
\d p_\theta&=&
{1\over r^{4}}\sprod{i\,\lbrac{\hat{\cal F}_\parallel}{\L_+}}{\d
X_+}\:.
\lb{ns34}
\end{eqnarray}
\addtolength{\jot}{-5pt}
We consider now the first variation of the YM equation (\ref{ns18}).
Its decomposition into real and imaginary parts yields
\addtolength{\jot}{5pt}
\begin{eqnarray}
-\,\frac{1}{NS^2}\,\d \ddot{X}_+&=&
-N\d X''_+
\,-\,\frac{(NS)'}{S}\d X'_+
\,-\,{i\over r^{2}}\lbrac{\L_+}        {\d\hat{\cal F}_\parallel\,}
\,+\,{i\over r^{2}}\lbrac{\hat{\cal F}_\parallel}{\d\L_+}
\nonumber\\
&&
\mbox{}\,-\,\d N\L''_+
\,-\,\d
\biggl(
{(NS)' \over S}
\biggr)
\L'_+
\lb{ns35}
\end{eqnarray}
and
\begin{eqnarray}
-\,\frac{1}{NS^2}\,\d \ddot{Y}_+&=&
-N
\biggl\{\,
\d Y''_+
\,+\,i\,\lbrak{\L_+}{\d\bcal{B}}'
\,+\, i\,\lbrak{\L_+'}{\d\bcal{B}}
\,\biggr\}
\nonumber\\
&&
\mbox{}\,-\,{(NS)' \over S}
\biggl\{\,
\d Y'_+
\,+\,i\,\lbrak{\L_+}{\d\bcal{B}}
\,\biggr\}
\,+\,\frac{i}{r^2}\,[\hat {\cal F}_\parallel,\d Y_+]\:.
\lb{ns36}
\end{eqnarray}
\addtolength{\jot}{-5pt}
The third term on the right hand side of (\ref{ns35}) is indeed real
and can be written, using (\ref{ns22}), as
\be
 \frac{i}{r^2}\,[\L_+,\d\hat {\cal F}_\parallel\,]=
\frac{1}{r^2}\,\ad(\L_+)\,\ad(\L_-)\,\d X_+\:.
\lb{ns37}
\ee
Equation (\ref{ns35}) can be simplified further. From (\ref{ns33}) and
the equilibrium equation (\ref{ns24}), we deduce
\addtolength{\jot}{5pt}
\begin{eqnarray*}
-\,\d N\L''_+&=&\frac{2}{r}\,\d m\,\L''_+\\
	     &=&\kappa N\Re\sprod{\L'_+}{\d X_+}\:\L''_+\\
	     &=&\kappa  \Re\sprod{\L'_+}{\d X_+}\:
		\biggl\{\,
		-{(NS)' \over S}\L_+
		\,+\,\frac{i}{r^2}\,[\hat {\cal F}_\parallel,\L_+]
		\,\biggr\}\:,
\end{eqnarray*}
\addtolength{\jot}{-5pt}
and the Einstein equations (\ref{ns11}), (\ref{ns12}) give
\be
-\,\d\,{(NS)' \over S}=-\frac{2}{r^2}\d m + \kappa r\,\d p_\theta\:.
\ee
If we use here  (\ref{ns33}) and (\ref{ns34}), we see that the last two
terms in  (\ref{ns35}) can be expressed as follows:
\addtolength{\jot}{5pt}
\begin{eqnarray}
&&\quad
\llap{$\displaystyle
-\,\d N\L''_+
$}
\,-\, \d\,{(NS)' \over S}
=\frac{1}{NS^2}
\Biggl\{\;
-{}\L_+'\:\kappa\, r\mu^2
\biggl\{
\frac{(NS)'}{S}+\frac{N}{r}
\biggr\}
\sprod{\L_+'}{\d X_+}
\quad\nonumber\\
&&
{}+\L_+'
\:\kappa\,\frac{\mu^2}{r}
\sprod{\,\lbrak{i\hat\bcal{F}_\parallel}{\L_+}}{\d X_+}
+\:\lbrak{i\hat\bcal{F}_\parallel}{\L_+}
\:\kappa\,\frac{\mu^2}{r}
\sprod{\L_+'}{\d X_+}
\:\:\Biggr\}\:\:,
\lb{ns38}
\end{eqnarray}
\addtolength{\jot}{-5pt}
where
\be
\mu^2:=\frac{NS^2}{r^2}\:.
\ee
Inserting these expressions into (\ref{ns35}) gives the following
pulsation equation for the real amplitude $\d X_+$ of the YM field:
\be
\d \ddot X_+ +\,U_{XX}\,\d X_+=0\:,
\lb{ns40}
\ee
where the operator $U_{XX}$ is given by
\addtolength{\jot}{8pt}
\begin{eqnarray}
U_{XX}&=&
{p_\ast}^2\,+\,\mu^2\ad(i\hat\bcal{F}_\parallel)
\,-\,\mu^2\ad(\L_+)\,\ad(\L_-)\nonumber\\
&&{}\:-\:\L_+'\:\kappa\,\mu^2
\Bigl\{
1-\kappa\, r^2 p_\theta
\Bigr\}
\sprod{\L_+'}{\cdot\,}\nonumber\\
&&{}\,+\:\L_+'
\:\kappa\,\frac{\mu^2}{r}
\sprod{\,\lbrak{i\hat\bcal{F}_\parallel}{\L_+}}{\cdot\,}
+\:\lbrak{i\hat\bcal{F}_\parallel}{\L_+}
\:\kappa\,\frac{\mu^2}{r}
\sprod{\L_+'}{\cdot\,}
\lb{ns41}\:,
\end{eqnarray}
\addtolength{\jot}{-8pt}
and $p_\ast$ denotes the differential operator
\be
p_\ast=-iNS{\displaystyle\frac{\partial}{\partial r}}\;.
\lb{ns39}
\ee

It is remarkable that the perturbations $\d Y_\pm$ and $\d\calb$ do not
appear in (\ref{ns40}) and that the back reaction of gravitation on $\d
X_+$ can be described by an effective potential (last three terms in
(\ref{ns41})).

Equation (\ref{ns36}) can easily be brought into the form
\be
\d \ddot Y_+ +\,U_{YY}\,\d Y_+ +\,U_{Y\calb}\sqrt{N}r\,\d \calb=0\:,
\lb{ns42}
\ee
where
\addtolength{\jot}{5pt}
\begin{eqnarray}
U_{YY}
&=&{p_\ast}^2\,+\,\mu^2\ad(i\hat\bcal{F}_\parallel)\:,\lb{ns43}\\
U_{Y\calb}&=&
p_\ast\,\mu\,\ad(\L_+)\,+\,\mu\,\ad(p_\ast\L_+)\:.\lb{ns44}
\end{eqnarray}
\addtolength{\jot}{-5pt}
We have thus achieved a partial decoupling, because neither $\d X_+$,
nor the metric perturbations, appear in (\ref{ns42}).

We proceed with the linearization of the YM equation (\ref{ns17}). The
variation of the last two terms is
\be
-\Bigl[\,\L_+\,,\,[\L_-,\d\bcal{B}]\,\Bigr]
\,+\,
[\L_+,\d\L'_-]\,-\,[\L'_-,\d\L_+]
\,+\,
\mbox{conjugate}\:,
\ee
which leads (with $\d\L_\pm=\d X_\pm \pm i\d Y_\pm$) to
\addtolength{\jot}{8pt}
\begin{eqnarray*}
&&-\,\biggl\{
\Bigl[\,\L_+\,,\,[\L_-,\d\bcal{B}]\,\Bigr]
+i\,[\L_+,\d Y'_-]+i\,[\L'_-,\d Y_+]
\:\biggr\}\\
&&
\qquad\hphantom{-\Bigl[\,\L_+\,,\,[\L_-,\d\bcal{B}]\,\Bigr]}
{}\,+\,\biggl\{\,
[\L_+,\d X'_-]-[\L'_-,\d X_+]\,\biggr\}
\,+\,
\mbox{conjugate}\:.
\end{eqnarray*}
\addtolength{\jot}{-8pt}
Here, the terms in the first curly bracket are in $LT$, while those in
the second are in $iLT$. The latter are compensated by their conjugates
and we find
\be
\sqrt{N}r\,\d \ddot \calb +\,U_{\calb\calb}\sqrt{N}r\,\d \calb
+\,U_{\calb Y}\,\d Y_+=0
\lb{ns45}
\ee
with
\begin{eqnarray}
U_{\calb\calb}&=& -\:\mu^2\ad(\L_+)\,\ad(\L_-)\:,\lb{ns46}\\
U_{\calb Y}&=&
-\:\mu\,\ad(\L_-)\,p_\ast\,+\,\mu\,\ad(\,p_\ast\L_-)\lb{ns47}\:.
\end{eqnarray}

At this point, we collect the results obtained so far as follows: Let
\be
\renewcommand{\arraystretch}{1.5}
\Phi =\left( \begin{array}{c}
\phi_Y \\
\phi_\bcal{B}
\end{array} \right)
=
\left( \begin{array}{c}
\d Y_+ \\
\sqrt{N}r\,\d\bcal{B}
\end{array} \right)\:,
\lb{ns48}
\renewcommand{\arraystretch}{1}
\ee
then (\ref{ns42}) and (\ref{ns45}) can be written as a $2\times 2$
matrix equation
\be
\ddot\Phi + U\Phi=0
\lb{ns49}
\ee
with
\be
\renewcommand{\arraystretch}{1.5}
U=\left( \begin{array}{cc}
     U_{YY} & U_{Y\calb} \\
U_{\calb Y} & U_{\calb\calb}
\end{array} \right)\:.
\lb{ns50}
\renewcommand{\arraystretch}{1}
\ee
The operators in this matrix are given in Eq. (\ref{ns43}),
(\ref{ns44}), (\ref{ns46}) and (\ref{ns47}).

The perturbation equations (\ref{ns40}) and (\ref{ns49}) do not include
the Gauss constraint (\ref{ns19}), whose linearization is easily found
to be
\be
\partial_t
\Bigl\{\,
p_\ast\frac{1}{\mu}\,\phi_\calb+\ad(\L_-)\,\phi_Y
\,\Bigl\}
=0\:.
\lb{ns51}
\ee
The role of this constraint will be discussed below.

In concluding this section, we emphasize once more, that the
perturbation equations hold also for black holes, if these are assumed
to be of essentially magnetic type (see Eq. (\ref{F1})). We also would
like to note that a comprehensive discussion of the pulsation equations
for the SU$(2)$ YM-Higgs sphaleron can be found in Ref.\ \cite{akiba}.
\section{Transformation to a hyperbolic system}
\setcounter{equation}{0}
\renewcommand{\theequation}{\arabic{section}.\arabic{equation}}
A look at the second order differential operator $U$ shows that it is
not elliptic and, thus, the system (\ref{ns49}) of partial differential
equations is not hyperbolic. With the help of the Gauss constraint
(\ref{ns51}) it is, however, possible to derive a hyperbolic system for
the subspace of physical perturbations orthogonal to a space of pure
gauge modes. This reformulation of the perturbation equations will turn
out to be very useful for several purposes.

We need first some notation. It is natural to introduce the following
scalar product for $LG_{\Bbb C}$-valued functions on
$(r_0,\infty)\subset{\Bbb R}_+$:
\be
\braket{\phi}{\psi}=\int_{r_0}^\infty\sprod{\phi}{\!\psi}\;\,d r_\ast
\lb{G1}
\ee
with the weighted measure
\[
d r_\ast=\frac{dr}{NS}\;.
\]
For a black hole, the lower limit $r_0$ is the radius of the horizon
and for a  regular solution it is zero. The operators $U_{XX}$ and $U$
are symmetric with respect to this scalar product on a dense domain of
$\bmrm{L}^2$-functions. This can be seen easily, using
\be
\braket{\phi}{p_\ast\psi}=\braket{p_\ast\phi}{\psi}
\lb{G2}
\ee
for smooth functions, which vanish at $r_0$, and
\be
\braket{\phi}{\ad (Z)\psi}=-\braket{\ad (c(Z))\phi}{\psi}
\lb{G3}
\ee
for arbitrary $LG_{\Bbb C}$-valued functions $\phi$, $\psi$, $Z$ in
$\bmrm{L}^2$ (see (\ref{ns10})).

A ``gauge mode'' $\Phi_G$ is  by definition a perturbation of the form
\be
\Phi_G=-\,i\calg\chi\;,
\lb{G4}
\ee
where $\calg$ is the linear operator
\be
\calg\chi=
\renewcommand{\arraystretch}{1.5}
\left( \begin{array}{c}
-\ad(\L_+)\chi\\
{\displaystyle\frac{1}{\mu}}\,p_\ast\chi
\end{array} \right)
\lb{G5}
\renewcommand{\arraystretch}{1}
\ee
and $\chi$ is a $\span{\S}_{\Bbb C}$-valued function. Note, that such
variations arise if (\ref{ns6}) is subjected to ($T$-valued) gauge
transformations $g=\exp(-\epsilon\chi)$. Eqn.\ (\ref{ns7}) and
(\ref{ns8}) show that this induces the infinitesimal transformation
\be
\L_+\rightarrow\L_+-\ad({\L_+}){\chi}\:,\qquad
\sqrt{N}r\,\calb \rightarrow
\sqrt{N}r\,\calb\,-i\,\frac{1}{\mu}\,p_\ast\chi\:.
\lb{G6}
\ee

It is not surprising that the following identity holds
\be
U\calg=0\;,
\lb{G7}
\ee
whence
\be
U\Phi_G=0\;.
\lb{G8}
\ee

``Physical perturbations'' $\Phi_P$ satisfy by definition
\be
\tilde{\calg}\Phi_P=0\;,
\lb{G9}
\ee
where $\tilde\calg$ is the linear operator
\be
\tilde{\calg}\Phi=p_\ast\frac{1}{\mu}\,\phi_\calb+\ad(\L_-)\,\phi_Y\;.
\lb{G10}
\ee
The component $\phi_Y$ is assumed to have values in the subspace
(\ref{E1}) of $LG_{\Bbb C}$ and $\phi_\calb$ has to be $\span{\S}_{\Bbb
C}$-valued.
Hence, physical perturbations are by definition those, for which the
curly bracket in (\ref{ns51}) vanishes.

Roughly speaking, a physical perturbation is orthogonal to all gauge
modes. More precisely, {\em modulo boundary terms} we have
\be
i\braket{\Phi_P}{\Phi_G}=\braket{\Phi_P}{\calg\chi}=
\braket{\tilde\calg\Phi_P}{\chi}=0\;,
\lb{G11}
\ee
which follows easily with Eq.\ (\ref{G2}) and (\ref{G3}).

The identity
\be
\tilde{\calg}U=0\;,
\lb{G12}
\ee
which can be verified by direct calculation, is related to the Gauss
constraint
\be
\partial_t\,\tilde\calg\Phi=0
\lb{G13}
\ee
in the following way: Assume Eq. (\ref{G13}) is satisfied for $t=t_0$,
then the dynamical equation (\ref{ns49}) implies that (\ref{G13}) is
satisfied for all times. Indeed, we conclude with (\ref{G12}) that
\be
{\partial_t}^2(\tilde\calg\Phi)
=\tilde\calg({\partial_t}^2\Phi)
=-\tilde\calg(U\Phi)=0\;.
\lb{G14}
\ee
As a corollary we have: A solution of (\ref{ns49}), which lies
initially in the physical subspace (\ref{G9}) and satisfies initially
the Gauss constraint (\ref{G13}), will satisfy the ``strong'' Gauss
constraint (\ref{G9}) for all times. For physical perturbations we can
thus use this strong form to bring Eq. (\ref{ns49}) to a hyperbolic
form. After some manipulations, one finds
\be
U=\Bigl\{{p_\ast}^2+V\Bigr\}-\calg\mu^2\tilde\calg\;,
\lb{G15}
\ee
where
\be
\renewcommand{\arraystretch}{1.5}
V=\left( \begin{array}{cc}
     V_{YY} & V_{Y\calb} \\
V_{\calb Y} & V_{\calb\calb}
\end{array} \right)
\lb{G16}
\renewcommand{\arraystretch}{1}
\ee
is the following (matrix-valued) potential
\addtolength{\jot}{5pt}
\begin{eqnarray}
V_{YY}        &=&\phantom{-}\mu^2 K^2
			  +\mu^2\ad(i\hat\bcal{F}_\parallel)\:,\lb{G17}\\
V_{Y\calb}
&=&\phantom{-}2(p_\ast\mu)K_++2\mu\,\ad(p_\ast\L_+)\:,\lb{G18}\\
V_{\calb Y}   &=&         -
2(p_\ast\mu)K_-+2\mu\,\ad(p_\ast\L_-)\:,\lb{G19}\\
V_{\calb\calb}&=&\phantom{-}\mu^2K^2
-\frac{({p_\ast}^2\mu)}{\mu}\:,\lb{G20}
\end{eqnarray}
\addtolength{\jot}{-5pt}
with
\begin{eqnarray}
K^2  &=&   -\ad(\L_+)\,\ad(\L_-)\:,\lb{G21}\\
K_\pm&=& \pm\ad(\L_\pm)\lb{G22}\:.
\end{eqnarray}

Modulo the strong Gauss constraint $\tilde\calg\Phi=0$,
Eq.\ (\ref{ns49}) is thus equivalent to
\be
{\partial_t}^2\Phi
=-\,\Bigl\{{p_\ast}^2+V\Bigr\}\Phi\;.
\lb{G23}
\ee
This system is clearly hyperbolic. We emphasize that this new system
implies the strong Gauss constraint for all times, if it is satisfied
initially:
$\tilde\calg\,\Phi|_{t_0}=\tilde\calg\,\partial_t\Phi|_{t_0}=0$. The
argument runs as follows: As a result of (\ref{G8}), (\ref{G15}) and
(\ref{G23}), $\tilde\calg\Phi$ satisfies the hyperbolic equation
\be
{\partial_t}^2(\tilde\calg\Phi)
=-\tilde\calg\calg\mu^2(\tilde\calg\Phi)
=-\Bigl\{p_\ast\frac{1}{\mu^2}\,p_\ast+K^2\Bigr\}\mu^2(\tilde\calg\Phi)\;.
\lb{G24}
\ee
Uniqueness of the Cauchy problem for the hyperbolic system (\ref{G24}),
with appropriate boundary conditions at $r_0$, then implies our claim.

We specialize now to harmonic perturbations proportional to
$\mbox{e}^{-i\omega t}$ and obtain for the amplitude of $\Phi$, denoted
by the same letter, the two eigenvalue problems
\be
U\Phi=\omega^2\Phi
\lb{G25}
\ee
and
\be
\Bigl\{{p_\ast}^2 +V\Bigr\}\Phi=\omega^2\Phi\:.
\lb{G26}
\ee
The second equation has the form of a (vector-valued) Schr\"odinger
equation.

In the next section, we prove that the spectrum of $U$ has a nonempty
negative part (which is presumably discrete), by constructing a smooth
trial function $\d\Phi$ for which $\bra{\d\Phi}\:U\,\ket{\d\Phi}$ is
strictly negative. This implies that the operator ${p_\ast}^2 +V$ has
also a negative part in the spectrum. This can be seen as follows:

If we can show that there exists a smooth function $\chi$, such that
\be
i\tilde\calg\d\Phi
=\tilde\calg\calg\chi
=\Bigl\{p_\ast\frac{1}{\mu^2}\,p_\ast+K^2\Bigr\}\chi\;,
\lb{G27}
\ee
then we have a decomposition
\be
\d\Phi=\d\Phi_P-\,i\calg\chi
\lb{G28}
\ee
into smooth physical and gauge components.
Using also (\ref{G8}), we have
\be
  \bra{\d\Phi}\:U\,\ket{\d\Phi}
= \bra{\d\Phi_P}\:U\,\ket{\d\Phi_P}
= \bra{\d\Phi_P}\:{p_\ast}^2 +V\ket{\d\Phi_P}
<0\:,
\lb{G29}
\ee
which would imply our claim.

Since $\tilde\calg\calg$ is a positive operator and since
$i\tilde\calg\d\Phi$ is smooth, we expect on the basis of elliptic
existence and regularity theorems that (\ref{G27}) has indeed a smooth
solution. This is one of several mathematical points which will be
discussed in the appendix. Another issue will be, whether the operator
${p_\ast}^2 +V$ is essentially self-adjoint on a dense domain of smooth
functions, which satisfy the boundary conditions implied by the physics
of the problem. This will be analyzed in section 6 and in the
appendix.

The relation between the operators $U$ and $Q:={p_\ast}^2+V$, given
explicitly in (\ref{G15}), can be summarized (on a formal level) as
follows: As a result of (\ref{G7}) and (\ref{G15}), both operators
split relative to the decomposition  of the $\mbox{L}^2$ space of
perturbations into physical and gauge degrees of freedom,
$\mbox{L}^2={\cal H}_P\oplus{\cal H}_G$, and their restrictions satisfy
\addtolength{\jot}{5pt}
\begin{eqnarray*}
 Q|_{{\cal H}_P}&=&U|_{{\cal H}_P}\;,\\
 U|_{{\cal H}_G}&=&0\;,\\
 Q|_{{\cal H}_G}&=&\calg\mu^2\tilde\calg|_{{\cal H}_G}\geq 0\;.
\end{eqnarray*}
\addtolength{\jot}{-5pt}
The last inequality follow from
\[
\bra{\Phi_G}{\,\calg\mu^2\tilde\calg}\ket{\Phi_G}
=
\bra{\tilde\calg\Phi_G}\,\mu^2\ket{\tilde\calg\Phi_G}
\lb{G31}
\]
for $\Phi_G\in{\cal H}_G$. In particular, the negative part of the
spectra of $U$ is contained in that of $Q$ and the discrete spectra of
the two operators coincide.
\section{Instability of generic EYM solutions}
\setcounter{equation}{0}
\renewcommand{\theequation}{\arabic{section}.\arabic{equation}}
We are now ready to establish the main point of this paper: For a given
solution with $\L_+=\sum_{\a\in\S}\, w_\a e_\a$, we construct a
one-parameter family of field configurations $\L_{(\tau)+}$,
${\calb}_{(\tau)}$ such that $\bra{\d\Phi}\:U\,\ket{\d\Phi}<0$ for the
variation
\be
\renewcommand{\arraystretch}{1.5}
\d\Phi =\left( \begin{array}{c}
{\d\phi}_Y \\
{\d\phi}_\bcal{B}
\end{array} \right)
=
\left( \begin{array}{c}
-i\,\partial_\tau\L_{(\tau)+}\,|_{\tau=0}\\
\sqrt{N}r\,\partial_\tau\,{\calb}_{(\tau)}\,|_{\tau=0}
\end{array} \right)\:.
\lb{H1}
\renewcommand{\arraystretch}{1}
\ee

The families we consider are of the form
\bea
\L_{(\tau)+}&=&\Ad(\exp(\tau Z))
\Bigl\{\,
\L_+\cos(\tau)+i\,T_+\sin(\tau)
\,\Bigr\}
\lb{H2}\:,\\
\calb_{(\tau)}&=&-\tau Z'\:,
\lb{H3}
\eea
where $T_+$ is a real element in the subspace (\ref{E1}), satisfying
\be
\lbrac{\,T_+}{T_-}=-2i\,\L_{3\parallel}\;,
\lb{H4}
\ee
and $Z$ is a $\span{\S}$-valued function of $r$ with
\be
\lim_{r\to r_0,\infty}\:\ad(\L_+)Z=i\,T_+\;,\qquad
\supp Z'\subset (r_0,\infty)\;.
\lb{H5}
\ee

If $\S$ is not empty such an element $T_+$ always exists (see appendix
A of ref.\ \cite{OB2}). A function $Z$ with the required properties can
be found if
\be
\lim_{r\to r_0,\infty}w_\a\neq 0\quad \mbox{for all }\:\:\a\in\S\;.
\lb{H6}
\ee
This can be seen as follows: Let $\{h_\a\}_{\a\in\S}$ be the dual basis
of $2\pi\S$ and put
\be
Z  =\sum_{\a\in\S}\, Z_\a\, h_\a\:,\qquad
T_+=\sum_{\a\in\S}\, T_\a\, e_\a
\lb{H7}
\ee
and
\be
Z_\a  =  \left \{
\begin{array}{ll}
-T_\a/w_\a (r_0)   &\mbox{\ for}\quad r<r_0 +(1-\epsilon)\\
-T_\a/w_\a (\infty)   &\mbox{\ for}\quad r>r_0 +(1+\epsilon)
\end{array} \right.
\lb{H8}
\ee
for an $\epsilon >0$. Then, both conditions in (\ref{H5}) are
satisfied.

For a regular (uncharged) solution, condition (\ref{H6}) is fulfilled
and $\S$ is not empty \cite{OB2}. Thus, a family (\ref{H2}), (\ref{H3})
{\em always\/} exists for solitons \cite{OB2}.

We note some properties of the families above. For the gauge group
SU$(2)$, these  are closely related to families studied by other
authors \cite{div}. The equilibrium solution is clearly obtained for
$\tau=0$. Applying a gauge transformation with $g=\exp(\tau Z)$, we
obtain
\be
\L_{(\tau)+}\to\L_+\cos(\tau)+i\,T_+\sin(\tau)\:,\qquad
\calb_{(\tau)}\to 0\:.
\lb{H9}
\ee
The first variations of (\ref{H2}) and (\ref{H3}) are
\be
\d\phi_Y=i\,\ad({\L_+})Z+T_+\:,\qquad
\d\phi_\calb=-i{\displaystyle\frac{1}{\mu}}\,p_\ast Z\:,
\lb{H10}
\ee
and these satisfy by construction the desired boundary conditions
\be
\lim_{r\to r_0,\infty}\d\phi_Y=0\:,\qquad
\lim_{r\to r_0,\infty}\d\phi_\calb=0\:.
\lb{H11}
\ee
($\d\phi_\calb$ has even compact support in $(r_0,\infty)$). Since an
equilibrium solution satisfies
\be
p_\ast\L_+|_{r_0}=p_\ast\L_+|_\infty=0\;,
\lb{H17}
\ee
we also have
\be
\lim_{r\to r_0,\infty}p_\ast\d\phi_Y=0\:.
\lb{H11a}
\ee

This choice of trial functions fulfills our goal: $\d\Phi$ is
normalizable and \\$\bra{\d\Phi}\:U\,\ket{\d\Phi}$ is finite and turns
out to be strictly negative.

The first of these two points is simple. Since $\d\phi_\calb$ in
(\ref{H10}) has compact support, we have to check only whether
\be
\int_{r_0}^\infty
\,|\d\phi_Y|^2\,\frac{dr}{NS}\:<\:\infty\:.
\lb{H12}
\ee
By construction,
\be
\d\phi_Y =
\left \{
\renewcommand{\arraystretch}{1.5}
\begin{array}{ll}
{\displaystyle\sum_{\a\in\S}}\;
T_\a\Bigl(w_\a/w_\a (r_0)-1\Bigr)\;e_\a
& \mbox{\ for}\qquad r<r_0 +(1-\epsilon)\:,\\
{\displaystyle\sum_{\a\in\S}}\;
T_\a\Bigl(w_\a/w_\a (r_\infty)-1\Bigr)\;e_\a
& \mbox{\ for}\qquad r>r_0 +(1-\epsilon)\:.\\
\end{array} \right.
\lb{Hn13}
\renewcommand{\arraystretch}{1}
\ee
Hence, the integrand has a finite limit for $r\rightarrow r_0$ (even
for extreme black hole solutions). Since $N$ and $S$ both approach one
at infinity, the integral is finite if $\L_+-\L_+(\infty)$ converges to
zero faster than $r^{-1/2}$.

The calculation of $\bra{\d\Phi}\:U\,\ket{\d\Phi}$ is somewhat tedious.
Considerable simplifications occur by separating a gauge mode in
$\d\Phi$:
\be
\d\Phi=\d\tilde\Phi  -i\calg Z
\lb{H14}
\ee
with
\be
\renewcommand{\arraystretch}{1.5}
\d\tilde\Phi =
\left(
\begin{array}{c}
T_+\\
0
\end{array}
\right)\:,\qquad
\calg Z=
\left(
\begin{array}{c}
-\,\ad(\L_+)Z\\
{\displaystyle\frac{1}{\mu}}\,p_\ast Z
\end{array}
\right)\:.
\lb{H15}
\renewcommand{\arraystretch}{1}
\ee
We stress, that neither $\d\tilde\Phi$ nor $\calg Z$ are normalizable.
Nevertheless, we have $U\calg Z=0$ and thus (\ref{H14}) and (\ref{H15})
give (with  a slight abuse of notation)
\begin{eqnarray}
\bra{\d\Phi}\:U\,\ket{\d\Phi}
&=&
 \bra{\d\tilde\Phi}\:U\,\ket{\d\tilde\Phi}
+i\braket{\calg Z}{U\d\tilde\Phi}
\nonumber\\
&=&
 \bra{\d\tilde\Phi}\:U\,\ket{\d\tilde\Phi}
+i\braket{U\calg Z}{\d\tilde\Phi}
+\sprod{\ad(p_\ast\L_+)Z}{T_+}\Bigm|_{r_0}^\infty
\nonumber\\
&=&
\bra{\d\tilde\Phi}\:U\,\ket{\d\tilde\Phi}\:.
\lb{H16}
\end{eqnarray}
The boundary term does not contribute because of Eq. (\ref{H17}). From
this, we obtain the intermediate result
\addtolength{\jot}{10pt}
\begin{eqnarray}
\bra{\d\Phi}\:U\,\ket{\d\Phi}
&=&
\phantom{2}
\int_{r_0}^\infty
\:\mu^2\sprod{T_+}{\ad(i\hat\bcal{F}_\parallel)\,T_+\,}\:dr_\ast
\nonumber\\
&=&
2\int_{r_0}^\infty
\:\mu^2\sprod{\hat\bcal{F}_\parallel}{\L_{3\parallel}}\:dr_\ast \:\;,
\lb{H18}
\end{eqnarray}
\addtolength{\jot}{-10pt}
where we have used (\ref{ns10}) and the property (\ref{H4}) of $T_+$.

Finally, we show, that the last term has a definite sign:
\be
2\int_{r_0}^\infty
\mu^2\sprod{\hat\bcal{F}_\parallel}{\L_{3\parallel}}\:dr_\ast
=
-\int_{r_0}^\infty
|\,p_\ast\L_{+}|^2
+2\mu^2|\hat {\cal F}_\parallel|^2
\:dr_\ast
\:.
\lb{H19}
\ee
After a partial integration, we find with the unperturbed YM equation
(\ref{ns18})
\be
\int_{r_0}^\infty
|\,p_\ast\L_{+}|^2
\:dr_\ast
=
-i\sprod{p_\ast\L_+}{\L_+}\Bigm|_{r_0}^\infty
-\int_{r_0}^\infty
\mu^2\sprod{\L_+}{\ad(i\hat{\cal F}_\parallel)\L_+}
\:dr_\ast \:.
\lb{H20}
\ee
The boundary term vanishes because of Eq. (\ref{H17}), and since
\begin{eqnarray}
2|\hat {\cal F}_\parallel|{\vphantom{|{\cal F}_\parallel}}^2
&=&\sprod{\hat {\cal
F}_\parallel}{i\,\lbrac{\L_+}{\L_-}-2\L_{3\parallel}}
\nonumber\\
&=&\sprod{\L_+}{\ad(i\hat{\cal F}_\parallel)\L_+}
-2\sprod{\hat {\cal F}_\parallel}{\L_{3\parallel}}\:,
\lb{H21)}
\end{eqnarray}
we have established the crucial result
\addtolength{\jot}{5pt}
\begin{eqnarray}
\bra{\d\Phi}\:U\,\ket{\d\Phi}
&=&
-\braket{p_\ast\L_{+}}{p_\ast\L_{+}}
-2\braket{\mu\hat {\cal F}_\parallel}{\mu\hat {\cal F}_\parallel}
\lb{Hns81}\nonumber\\
&=&
-\int_{r_0}^\infty
\Bigl\{
N\,|\L_{+}'|{\vphantom{|\L_+|}}^2
+\frac{2}{\,r^2}|\hat {\cal F}_\parallel|{\vphantom{|{\cal
F}\_parallel}}^2
\Bigr\}S\;dr\:.
\lb{H22}
\end{eqnarray}
\addtolength{\jot}{5pt}
This expression is clearly finite and strictly negative.

One can show, that expression (\ref{H22}) is also equal to the second
variation of the Schwarzschild mass for the one-parameter family
(\ref{H2}), (\ref{H3}). (This is the way we arrived originally at the
variation (\ref{H10})). For a systematic discussion of the relation
between variational principles for the spectra of radial pulsations and
second variations of the total mass, we refer to \cite{OB4}.

In summary, we have proven (apart from technical subtleties) that
static, spherically symmetric, asymptotically flat solutions of the EYM
equations are unstable. More precisely, we have established:
\begin{theorem} A generic, regular solution is {\em unstable}, if the
(magnetic) YM charge vanishes (i.e., if $\,\lim_{r\to\infty}\L(r)$ is a
homomorphism from $L${\rm SU}$(2)$ to $LG$) and if asymptotically
$\L_+-\L_+(\infty)\sim r^{-\a}$ with $\a>1/2$.
\end{theorem}

For a black hole (with horizon at $r_h$ and $\L_+=\sum_{\a\in\S}\, w_\a
e_\a$), the assumptions are somewhat more restrictive and ``trivial''
solutions have to be excluded. We call a generic, essentially magnetic
solution ``trivial'' if either $\S$ is empty or each amplitude
$\omega_\alpha$ is constant. These are clearly just the Reissner-Nordstr\"om
solutions.
\begin{theorem} A generic, essentially magnetic, non-trivial black hole
solution is {\em unstable}, if $\,\lim_{r\to r_h,\infty}w_\a\neq 0$ for
all $\a\in\S$ and if asymptotically $\L_+-\L_+(\infty)\sim r^{-\a}$
with $\a>1/2$.
\end{theorem}

We would like to stress that we were able to draw this conclusion,
assuming only weak asymptotic conditions for the solutions. In
particular, the fall-off condition is mild and is certainly fulfilled
for the Bartnik-McKinnon and the related black hole solutions, as was
shown rigorously in \cite{maison}. The same is true for the regular
solutions, which have been found numerically by H.P.\ K\"unzle for the
group SU$(3)$ \cite{kunzle2}. (For both types, the exponent $\alpha$ is
equal to one.)
\section{Sphaleron-like instabilities as bound
states of a fictitious deuteron problem}
\setcounter{equation}{0}
\renewcommand{\theequation}{\arabic{section}.\arabic{equation}}
We address now the question, whether the operator $p_\ast^2+V$ in the
eigenvalue problem (\ref{G26}) is essentially self-adjoint on a dense
domain of smooth functions, which satisfy the boundary conditions
implied by the physics of the problem. That this is indeed the case,
will be shown in the present section for SU$(2)$ solitons. The
discussion of the general case is deferred to the appendix.

For regular SU$(2)$ solutions, it turns out, that the eigenvalue
equation (\ref{G26}) can be interpreted as a fictitious deuteron
problem for a neutron-proton potential, consisting of a central part, a
tensor force and a spin-orbit coupling. All parts are determined by the
unperturbed soliton and can be shown to be bounded. The corresponding
Schr\"odinger operator is thus essentially self-adjoint on the subspace
of smooth functions with compact support and self-adjoint on the
Sobolev space $\mbox{H}^2({\Bbb R}^3)$. These facts will be used later
in an analysis of the instabilities, implied by the existence of bound
states (see the appendix).

In order to bring the operator $p_\ast^2+V$ to a standard Schr\"odinger
form, we introduce the new radial coordinate
\be
\rho(r)=\int_0^r\frac{d\,y}{NS}
\lb{I1}
\ee
in terms of which $p_\ast=-id/d\rho$. Since $\mu^2$ behaves like
$1/\rho^2$ near the origin, we separate from the potential (\ref{G16})
the singular term
\be
V(\rho)=\frac{J^2}{\rho^2}+\tilde V(\rho)\;,
\lb{I2}
\ee
where
\be
\renewcommand{\arraystretch}{1.5}
J^2=\left( \begin{array}{cc}
     K^2(0) & 2iK_+(0) \\
  -2iK_-(0) & K^2(0)+2
\end{array} \right)
\lb{I3}
\renewcommand{\arraystretch}{1}
\ee
and the remainder $\tilde V$ is bounded.

For a generic soliton, the eigenvalues of $J^2$ are equal to $j(j+1)$
with $j=k\pm 1$, whereby the integer $k$ runs through a (strictly)
positive, finite set. This set always contains $k=1$ and is uniquely
determined by $\L_3=\L_{3\parallel}$.

In a representation in which $J^2$ is diagonal, $J^2/\rho^2$ thus
describes the central barriers of a finite set of partial waves.

We now discuss in detail the equations for the gauge group SU$(2)$. For
this group, only S and D waves occur in Eq. (\ref{G26}). In the
variable $\rho$ and with the parametrization
\begin{eqnarray}
\L_+&=&w\,\tau_+ = w\,(\tau_1+i\tau_2)\;,\nonumber\\
\L_3  &=&\tau_3\nonumber
\end{eqnarray}
and
\be
\renewcommand{\arraystretch}{1.5}
\left( \begin{array}{c}
\phi_Y \\
\phi_\bcal{B}
\end{array} \right)
 =
\frac{u_S}{\sqrt{3}}
\left( \begin{array}{c}
\tau_+ \\
\tau_3
\end{array} \right)
+\,
\frac{u_D}{\sqrt{3}}
\left( \begin{array}{c}
-\tau_+ \\
\tau_3
\end{array} \right)\:\;,
\lb{I4}
\renewcommand{\arraystretch}{1}
\ee
the eigenvalue equation (\ref{G26}) takes the form
\be
\renewcommand{\arraystretch}{1.5}
\Biggl\{
-\frac{d^2}{d\rho^2}
+
\frac{1}{\rho^2}
\left( \begin{array}{cc}
     0 & 0 \\
     0 & 6
\end{array} \right)
+
\tilde{V}
\;\Biggr\}
\left( \begin{array}{c}
u_S \\
u_D
\end{array} \right)
=
\omega^2
\left( \begin{array}{c}
u_S \\
u_D
\end{array} \right)
\lb{I5}
\renewcommand{\arraystretch}{1}\:\;.
\ee
For the potential $\tilde V$ we find
\be
\renewcommand{\arraystretch}{1.5}
\tilde V=\frac{1}{3}\left( \begin{array}{cc}
\tilde V_{SS}     & \tilde V_{SD} \\
\tilde V_{DS}& \tilde V_{DD}
\end{array} \right)\:\;,
\lb{I6}
\renewcommand{\arraystretch}{1}
\ee
where
%
\begin{eqnarray}
\tilde V_{SS} &=&
\Bigl\{\mu''/\mu+8(\mu w)'+6\mu^2w^2\Bigr\}
+2\mu^2(1-w^2)
\;,\lb{I7}\\
\tilde V_{SD} &=&
\sqrt{2}\Bigl\{\mu''/\mu+2(\mu w)'\Bigr\}
-\sqrt{2}\mu^2(1-w^2)
\;,\lb{I8}\\
\tilde V_{DS} &=& \tilde V_{SD}\lb{I9}\;,\\
\tilde V_{DD} &=&
2\Bigl\{\mu''/\mu-4(\mu w)'+3\mu^2w^2-9/\rho^2\Bigr\}
+\mu^2(1-w^2)
\lb{I10}
\end{eqnarray}
%
and a {\em dash} denotes a derivative {\em with respect to} $\rho$.

It is amusing and helpful to note, that this coupled eigenvalue problem
has the same form as the Schr\"odinger equation for the relative motion
of a two-body proton-neutron system with the three standard terms
$V_C(r)$ (central potential), $V_T(r)\,\mbox{\bf S}_{12}$ (tensor
interaction) and $V_{LS}(r)\,\mbox{\bf L$\cdot$S}$ (spin-orbit
interaction). For total angular momentum $J=1$ and total spin $S=1$,
the possible orbital angular momenta are $L=1$ and $L=0,2$. Because of
parity conservation, the P wave decouples from the S and D waves. The
remaining equation, describing coupled S and D waves, reads in suitable
units
\begin{eqnarray}
\renewcommand{\arraystretch}{1.5}
&&\Biggl\{
-\frac{d^2}{dr^2}
+
\frac{1}{r^2}
\left( \begin{array}{cc}
     0 & 0 \\
     0 & 6
\end{array} \right)
\nonumber\\
&&\qquad
+
\left( \begin{array}{cc}
V_C          & \sqrt{8} V_T \\
\sqrt{8} V_T & V_C-2V_T-3V_{LS}
\end{array} \right)
\;\Biggr\}
\left( \begin{array}{c}
u_S \\
u_D
\end{array} \right)
=
E
\left( \begin{array}{c}
u_S \\
u_D
\end{array} \right)
\lb{I11}
\renewcommand{\arraystretch}{1}\:\;.\qquad
\end{eqnarray}
These equations have first been derived by Rarita and Schwinger
\cite{rarita}. Our eigenvalue problem (\ref{I5}) is clearly just a
special case of (\ref{I11}) and we can, by identification, express the
three potentials in terms of the functions $N,S,w$ of the
Bartnik-McKinnon solutions.

We present numerical results elsewhere (see also Ref.\ \cite{maison2})
and emphasize here only, that with this interpretation the mathematical
nature of our eigenvalue problem is automatically settled, because the
perturbation $\tilde V$ is completely harmless. We come back to this in
the appendix, where we discuss also the operator corresponding to the
strong Gauss constraint.
\section*{Acknowledgments}
Special thanks go to Dieter Maison for very useful comments.
Interesting discussions with members of our theory group, especially
with George Lavrelashvili and Michael Volkov, are gratefully
acknowledged. We would also like to thank Markus Heusler for
discussions at an earlier stage of our work. Finally, we wish to thank
the Swiss National Science Foundation for financial support.

\vskip 2.5\baselineskip
\noindent{\Large\bf Appendix}
\appendix
\vskip \baselineskip
\setcounter{equation}{0}
\renewcommand{\theequation}{\Alph{section}\arabic{equation}}
In the main body of the text, we deferred on several occasions some of
the mathematical subtleties to this appendix.
\section{Essential self-adjointness of the effective\protect\\
hamiltonian}
For black holes, the operator $Q=p_\ast^2+V$ in (\ref{G15}), with the
expressions (\ref{G17}) -- (\ref{G20}) for the matrix-valued potential
$V$, is effectively a standard Schr\"odinger operator on the whole real
line (see \cite{NS3}) and is thus essentially self-adjoint on
$C^\infty$ functions with compact support. (The potential $V$ is
bounded for black holes.) For solitons, we can use Weyl's limit point
-- limit circle criterion (see \cite{reed}, Sec.\ X.1 or
\cite{weidmann}) for the first two terms of the operator
\be
Q=-\frac{d^2}{d\rho^2}+\frac{J^2}{\rho^2}+\tilde V(\rho)
\lb{app1}
\ee
(see (\ref{I2}), (\ref{I3})). Since $\tilde V$ is bounded, the
Rellich-Kato theorem implies, that the domains of (essential)
self-adjointness are not changed by this additive term.

Another method which will be used later, is to lift $Q$ to a
Schr\"odinger operator $H_Q$ on ${\Bbb R}^3$ and to use powerful
results for this kind of operators. In Sec.\ 6 we showed how this can
be achieved, if the gauge group is SU$(2)$: $H_Q$ can then be chosen to
be of the standard form for a deuteron problem. This operator is
essentially self-adjoint on $\mbox{C}_0^2({\Bbb R}^3)\otimes{\Bbb
C}^{\,4}$ and self-adjoint on the Sobolev space $\mbox{H}^2({\Bbb
R}^3)\otimes{\Bbb C}^{\,4}$ (see, e.g., \cite{reed}, Sec.\ X.2).
Restricting these domains to the subspace of S and D waves, provides
the domains we are interested in for the original operator $Q$. For
instance, $Q$ is essentially self-adjoint on
\be
{\cal D}(Q)=
\Bigl\{\,
(u_S,u_D) \:\Big|\:
u_S\in\mbox{C}_0^\infty[0,\infty),u_S(0)=0;
u_D\in\mbox{C}_0^\infty(0,\infty)
\,\Bigr\}\:.
\lb{app2}
\ee

Although, we have not yet generalized this construction to arbitrary
gauge groups, the generalization of ${\cal D}(Q)$ is obvious: The S
waves have to be restricted as in (\ref{app2}) and the higher waves
have to lie in $\mbox{C}_0^\infty(0,\infty)$. We also note at this
point that the variation (\ref{H10}) lies in the domain of definition
of the self-adjoint extension of $(Q,{\cal D}(Q))$.

If we would restrict the S waves also to $\mbox{C}_0^\infty(0,\infty)$,
the operator $Q$ would not be essentially self-adjoint. For each S wave
sector, it would actually have a one-parameter family of self-adjoint
extensions. The self-adjoint extension, given above, is just the
Friedrichs extension, and one can show that it is the only one which is
compatible with the strong Gauss constraint (\ref{G9}).

The existence and smoothness problems in connection with
Eq.\ (\ref{G27}) can also be solved by lifting the equation to ${\Bbb
R}^3$ and using standard existence and regularity theorems for elliptic
operators on ${\Bbb R}^3$. (The details can easily be worked out for
$G=\mbox{SU}(2)$.)
\section{Spectral properties and unstable perturbations}
\setcounter{equation}{0}
\renewcommand{\theequation}{\Alph{section}\arabic{equation}}
In Sec.\ 4 it was shown that the perturbation equations for even parity
perturbations are equivalent to the hyperbolic system
\be
{\partial_t}^2\Phi
=-Q\,\Phi\;.
\lb{app3}
\ee
We recall also that these equations imply the propagation of the strong
Gauss constraint. As a main point of this paper we proved that the
self-adjoint operator $Q$, restricted to the subspace of physical
states, satisfying the strong Gauss constraint, has a non-empty
negative spectral part. This fact implies, of course, that there are
unstable Hilbert space solutions of (\ref{app3}). We just have to
choose the initial data $\Phi_0$ such that
$\mbox{E}_Q(-\infty,0)\,\Phi_0\neq0$, where $\mbox{E}_Q(\cdot)$ denotes
the projection valued measure belonging to $Q$ (see below). It is even
possible to choose $\Phi_0\in\mbox{C}_0^\infty$, because the smooth
functions with compact support are dense in the Hilbert space
$\mbox{L}^2$.

The question now arises, whether such a Hilbert space solution is even
a (classical) solution of the system of partial differential equations
(\ref{app3}), in other words, whether the Hilbert space solutions with
$\Phi_0\in\mbox{C}_0^\infty$ are automatically smooth. For black holes,
the positive answer to this question is contained in a paper by Wald
\cite{wald2}. His analysis does, however, not directly apply to
solitons, because he assumed, that space is a {\em complete\/}
Riemannian manifold.

A direct attack of the problem on the half-line $(0,\infty)$ is again
difficult. Once more, a way out is lifting the problem to ${\Bbb R}^3$,
\be
{\partial_t}^2\Phi
=-H_Q\,\Phi\;,
\lb{app4}
\ee
where the analysis of \cite{wald2} applies. We showed earlier, how this
can be done for SU$(2)$. Since the required smoothness properties
certainly do not depend on the gauge group, it is not worthwhile to
elaborate further on this. We would like, however, to present here a
simplified version of Wald's argument.

Consider a hyperbolic system on ${\Bbb R}\times{\Bbb R}^n$ of the form
(\ref{app4}) with a smooth elliptic operator $A$ (instead of $H_Q$),
which is essentially self-adjoint on $\mbox{C}_0^\infty({\Bbb R}^n)$.
For systems of this kind, a lot is known about the Cauchy problem (a
standard reference is \cite{horm}). In particular, one knows (see
Theorem 23.2.2 in Vol.\ III of \cite{horm}), that the Cauchy problem
with initial data $\Phi_0,\dot\Phi_0\in\mbox{C}_0^\infty({\Bbb R}^n)$
has a unique solution in $\mbox{C}^\infty({\Bbb R}\times{\Bbb R}^n)$,
which for any fixed time $t$ is in $\mbox{C}_0^\infty({\Bbb R}^n)$.
This smooth solution must agree with the Hilbert space solution of the
Cauchy problem because the latter is also unique.

Let us now assume that the spectrum $\sigma(A)$ of $A$ has a non-empty
intersection $\sigma(A)_-$ with $(-\infty,0)$. We also assume that
$\sigma(A)$ is bounded from below (this is the case for $A=H_Q$). The
Hilbert space solution of the Cauchy problem can easily be expressed in
terms of the projection valued measure $\mbox{E}(\cdot)$ belonging to
$A$. It suffices, for what follows, to take as initial data
$\Phi|_{t=0}=\Phi_0,\partial_t\Phi|_{t=0}=0$. Then, the corresponding
Hilbert space solution of
\be
\ddot\Phi_t
=-A\,\Phi_t
\lb{app5}
\ee
is
\addtolength{\jot}{5pt}
\begin{eqnarray}
\Phi_t\: =\:
\mbox{E}(\{0\})\Phi_0&\!+\!&
\int_{(0,\infty)}\cos(t\sqrt{\l})\:\:d\,\mbox{E}(\l)\Phi_0
\nonumber\\
&\!+\!&\int_{\s(A)_-}\cosh(t\sqrt{-\l})\:\:d\,\mbox{E}(\l)\Phi_0\:,
\lb{app6}
\end{eqnarray}
\addtolength{\jot}{-5pt}
as can easily be verified. Note in particular, that $\Phi_{t=0}=
\mbox{E}({\Bbb R})\Phi_0=\Phi_0$, as required.

With standard rules (see, e.g., \cite{rudin} Chap.\ 13), we obtain from
this
\addtolength{\jot}{5pt}
\begin{eqnarray}
\braket{\Phi_0}{\Phi_t}
\:=\:\|\mbox{E}(\{0\})\Phi_0\|^2
&\!+\!&
\int_{(0,\infty)}\cos(t\sqrt{\l})\:\:d\mu_{\Phi_0}(\l)
\nonumber\\
&\!+\!&\int_{\s(A)_-}\cosh(t\sqrt{-\l})\:\:d\mu_{\Phi_0}(\l)
\lb{app7}
\end{eqnarray}
\addtolength{\jot}{-5pt}
and
\be
\|\Phi_t\|^2\geq
\int_{\s(A)_-}\cosh(t\sqrt{-\l})\:\:d\mu_{\Phi_0}(\l)\;,
\lb{app8}
\ee
where $\mu_{\Phi_0}$ is the finite measure
$\braket{\Phi_0}{\mbox{E}(\cdot)\Phi_0}$ on $\Bbb R$, whose support is
contained in $\s(A)$. As emphasized above, we can choose $\Phi_0$ such
that  $(\supp\mu_{\Phi_0})\cap\s(A)_-$ is non-empty. Then (\ref{app7})
and (\ref{app8}) imply that both quantities on the left diverge
exponentially. This exponential grows translates to an average
exponential grows of the {\em classical\/} solution of the hyperbolic
system for smooth initial data with compact support.

These considerations conclude our instability proof.

\end{document}